\documentclass[preprint2]{aastex}

\shorttitle{Density modulation index in the inner solar wind}
\shortauthors{SK Bisoi et al.}

\begin{document}

\title{A study of density modulation index in the inner heliospheric solar 
wind during solar cycle 23}

\author{Susanta Kumar Bisoi and P. Janardhan}
\affil{Astronomy \& Astrophysics Division, Physical Research Laboratory, Ahmedabad 380 009, 
India.}    
\email{susanta@prl.res.in,jerry@prl.res.in}
\and
\author{M. Ingale and P. Subramanian}
\affil{Indian Institute of Science Education and Research, Dr. Homi Bhabha Road, Pashan, Pune 
411 021, India.}
\email{i.madhusudan@students.iiserpune.ac.in,p.subramanian@iiserpune.ac.in}
\and
\author{S. Ananthakrishnan}
\affil{Department of Electronic Science, University of Pune, Pune 411 007, India.}    
\email{subra.anan@gmail.com}
\and
\author{M. Tokumaru, and K. Fujiki}
\affil{Solar-Terrestrial Environment Laboratory, Nagoya University, Nagoya 464-8601, Japan.} 
\email{tokumaru@stelab.nagoya-u.ac.jp, fujiki@stelab.nagoya-u.ac.jp}
\begin{abstract}
The ratio of the rms electron density fluctuations to the background density in the solar wind 
(density modulation index, $\epsilon_{N} \equiv \Delta{N}/N$) is of vital importance in 
understanding several problems in heliospheric physics related to solar wind turbulence.  In 
this paper, we have investigated the behavior of $\epsilon_{N}$ in the inner-heliosphere from 
0.26 to 0.82 AU.  The density fluctuations $\Delta{N}$ have been deduced using extensive 
ground-based observations of interplanetary scintillation (IPS) at 327 MHz, which probe spatial 
scales of a few hundred km.  The background densities ($N$) have been derived using
near-Earth observations from the Advanced Composition Explorer ({\it{ACE}}).  Our analysis 
reveals that $0.001 \lesssim \epsilon_{N} \lesssim 0.02$ and does not vary appreciably with 
heliocentric distance.  We also find that $\epsilon_{N}$ declines by 8\% from 1998 to 2008.  
We discuss the impact of these findings on problems ranging from our understanding of Forbush 
decreases to the behavior of the solar wind dynamic pressure over the recent peculiar solar 
minimum at the end of cycle 23.
\end{abstract}


\section{Introduction}
 \label{S-Intro}
The solar wind is an unparalleled natural laboratory for the study of magneto-hydrodynamic 
turbulence e.g., \citep{TMa95,GRM95,BCa05,Mar06,Spa09}. It involves fluctuations in 
magnetic field, density and velocity over a wide range of spatial and temporal scales.  
Turbulent density fluctuations in the solar wind have been observed over heliocentric 
distances ranging from $\sim$0.14 AU or 30$R_{\odot}$ to 1 AU or 215 $R_{\odot}$ 
from the Sun, where R${_{\odot}}$ is the solar radius 
\citep{Col78,MTu90,BBr95,JaB96,EfR00,Spa02,BiJ03,Spa09,TKF12}.  Moreover, density 
fluctuations are often believed to be better tracers of solar wind flows as compared to 
solar wind density \citep{ACK80,WoA95,HWN95}.  Detailed measurements of solar wind 
density fluctuations near the Earth have been made using {\it{in-situ}} data from 
spacecraft, such as {\it{Helios 1}}, {\it{Helios 2}}, {\it{Wind}}, and {\it{Ulysses}}. 

MHD turbulence theory generally assumes incompressibility, and density fluctuations do 
not fit into the narrative.  Furthermore, the scaling law in (spatial) wavenumber space 
exhibited by density turbulence observations is generally consistent with the Kolmogorov 
theory, which in fact holds for incompressible fluid turbulence in the absence of magnetic 
fields.  The implications of compressibility (as evidenced by observations of turbulent 
density fluctuations) vis-a-vis theories of MHD turbulence is a subject of considerable 
discussion \citep{TMa94,HCR05,SZa10}.  In particular, knowing the manner in which the 
density modulation index 
\begin{equation}
\epsilon _{N} \equiv \frac{\Delta N}{N}
\label{eqmodindex}
\end{equation}
varies with distance from the Sun is of vital importance for a variety of applications.
 
In the expression for $\epsilon_{N}$ (Eq~\ref{eqmodindex}), the quantity $\Delta N$ represents 
the turbulent density fluctuation while $N$ is the background density.  An understanding of 
$\epsilon_{N}$ is important for understanding turbulent dissipation and consequent local heating 
of the solar wind \citep{CaM09}.  It is also an important ingredient in constructing models 
for the quantity $C_{N}^{2}$, which is the amplitude of the density turbulence spectrum 
\citep{TMa08}.  In turn, $C_{N}^{2}$ is crucial in understanding angular broadening of radio 
sources due to solar wind turbulence \citep{JaL93,Bas94,SCa11} and in explaining the rather 
low brightness temperatures of the solar corona at meter to decameter wavelengths \citep{TMa08}.  
A crucial role is also played by $\epsilon_{N}$ in influencing the propagation of energetic 
electrons, produced by solar flares and other explosive solar surface phenomena, through the 
heliosphere \citep{RKo10}. 

Recently, using IPS measurements of scintillation index from 1983 to 2009, the solar wind 
micro-turbulence levels in the inner heliosphere were shown to be steadily declining since 
$\approx$1995 \citep{JaB11}.   Using ground-based magnetograms from the National Solar 
Observatory at Kitt Peak (NSO/KP) a steady and systematic decline in solar polar fields, 
starting from $\approx$1995, has also been reported \citep{JBG10,BiJ14}.  In addition, 
both in-ecliptic ({\it{ACE}} and {\it{Wind}}) \citep{JRL11} and out-of-ecliptic ({\it{Ulysses}}) 
\citep{McE08} solar wind measurements, during the recent minimum of solar cycle 23, in 
2008\,--\,2009, have shown a reduction in solar wind dynamic pressure of about 20\%.  Under 
these very unusual and unique circumstances of declining solar polar field strengths and 
density turbulence levels ($\propto\Delta$N) \citep{JBG10,JaB11,BiJ14}, studies of the 
temporal changes of $\epsilon_{N}$ in the inner-heliosphere are both important and 
crucial in understanding the relation between magnetic field fluctuations and density 
fluctuations.  Such a study also impinges on the important question of the role 
of the dynamic pressure exerted by the solar wind on the Earth's magnetosphere during this 
unusual phase.

The first measurements of $\epsilon_{N}$ were made at heliocentric distances $\lesssim$ 40~$R_{\odot}$,
by \cite{WoA95} using {\it{Ulysses}} measurements obtained in 1991.  Subsequently, density fluctuations 
in different types of solar wind flows have been reported at 1 AU \citep{HWN95} and also in the region 
from 0.3 to 1 AU using the {\it{Helios 2}} spacecraft, interplanetary plasma data, obtained with a time 
cadence of 45 mins \citep{BBr95}.  These authors reported a $\epsilon_{N}$ of $\approx$ 0.1 and proposed 
that compressive phenomena were not strong enough at the 45 minute cadence used for the observations.  
Further, \cite{Spa02} reported a $0.06 \lesssim \epsilon_{N} \lesssim 0.15$ in the heliocentric distance 
range 16\,--\,26~$R_{\odot}$.  Using {\it{Wind}} spacecraft data at 1 AU, \cite{SSp04} have estimated 
$\epsilon_{N}$ of the order of 0.03\,--\,0.08 and proposed both a linear and quadratic relationship between 
the $\epsilon_{N}$ and the magnetic field index ($\epsilon_{B}$) in regions of the near-sun solar wind.  
The data used in previous papers have been sparse, with either the observations being confined to a small 
region of the heliosphere or covering periods from a few days to years.  However, in this paper, we have 
made use of observations spanning the whole inner-heliosphere covering the heliocentric distance range of 
0.26\,--\,0.82 AU corresponding to 55\,--\,175 {$R_{\odot}$}.  In addition, our data set of eleven years 
covers the whole of solar cycle 23, thereby enabling a study of the long term temporal variation in 
$\epsilon_{N}$ as well.  

In this study, we have made use of extensive and systematic IPS measurements to investigate the radial 
evolution of $\epsilon_{N}$ defined in Eq~(\ref{eqmodindex}).  While electron density fluctuations have 
been estimated at 327 MHz using measurements from the multi-station IPS observatory of the Solar-Terrestrial 
Environment Laboratory (STEL), Japan, the solar wind densities used were derived 
from {\it{in-situ}} observations from the {\it{ACE}} spacecraft \citep{StF98} with $\epsilon_{N}$ 
being estimated for the period 1998\,--\,2008, covering the whole of solar cycle 23.  

The rest of the paper is organized as follows: section \ref{S-Ips} briefly discusses 
interplanetary scintillation as well as phase modulation of plane waves by the solar 
wind.  In section \ref{S-Data}, the use of IPS and {\it{ACE}} data and their analyses are discussed.  
Subsequently, in section \ref{S-mod-index} we verify the long term temporal and spatial behavior of 
$\epsilon_{N}$.  Finally, section \ref{S-dis} summarizes our results. 

\section{Interplanetary scintillation}
 \label{S-Ips}
\begin{figure}[ht]
\vspace{4.9cm}
\includegraphics{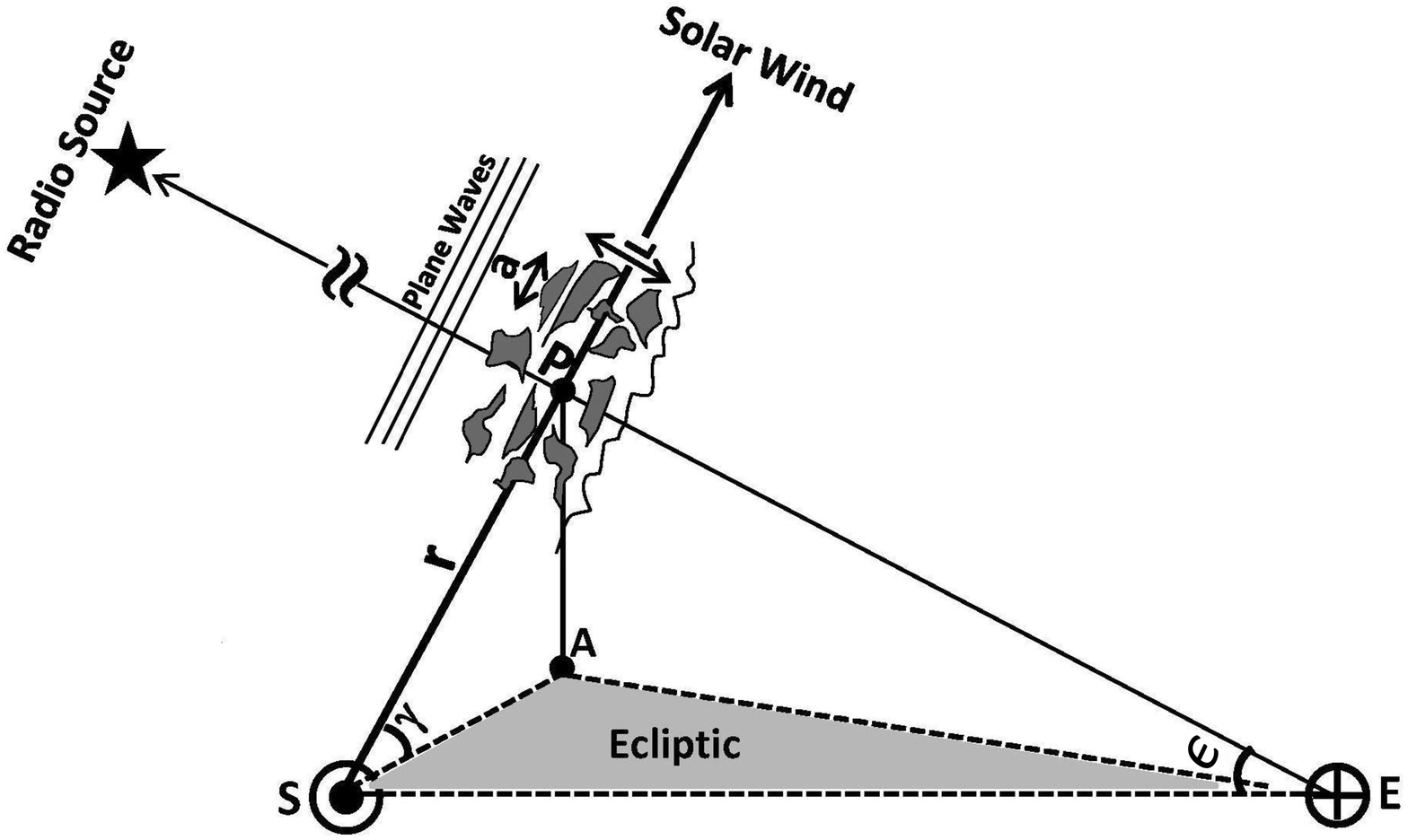}
\caption{ A schematic of the IPS observing geometry. The Earth, the Sun, the point 
of closest approach of the LOS to the Sun, and the foot point of a perpendicular from P 
to the ecliptic plane are shown by points E, S, P and A while the angles $\epsilon$ and 
$\gamma$ are the solar elongation and heliographic latitude of the observed source.}
\label{Fig1}
\end{figure}

IPS is a diffraction phenomenon in which coherent electromagnetic radiation 
from a distant radio source passes through the turbulent and refracting solar 
wind and suffers scattering.  This results in random temporal variations of 
the signal intensity (scintillation) at the Earth.  A schematic of the typical 
IPS  observing geometry is shown in Figure \ref{Fig1}. The broken lines in 
Figure \ref{Fig1} lie in the ecliptic plane, while the solid lines lie out of 
the ecliptic plane.  The long-dashed line is the orbit of the Earth around the 
Sun.  The line-of-sight (LOS) to a distant compact radio source with respect 
to the Sun (`S') and the Earth (`E') is shown by a solid line from E passing 
through the point `P', the point of closest approach of the LOS to the Sun.
%
\protect	  
%
The angles $\epsilon$ and $\gamma$ are respectively, the solar elongation and 
heliographic latitude of the source while `A' is the foot point of a perpendicular 
from P to the ecliptic plane. The heliocentric distance `r' of the radio source, 
in AU, is given by r = sin($\epsilon$).  It must be noted that the scintillations 
observed at the Earth are modulated by the Fresnel filter function
$\mathsf{Sin^{2}(\frac{q^{2}\lambda z}{4\pi})}$ where, q is the wave number of the 
irregularities, z is the distance from E to P, and $\lambda$ is the observing 
wavelength. Due to the action of the Fresnel filter, IPS observations at 327 MHz 
enable one to probe solar wind electron density fluctuations of scale sizes 
$\leq$1000 km both in and out of the ecliptic \citep{RBA74,CFi85,YaT98,FBD08} and 
over a wide range of heliocentric distances in the inner-heliosphere \citep{JaB96}.  

Besides density fluctuations of spatial scale sizes $\leq$ $10^{3}$ km, there 
are largerscale solar wind density fluctuations caused by structures such as 
coronal mass ejections (CMEs) and solar flares, which originate on the solar surface.  
The typical scale sizes of these structures range from $10^{4}$ to $10^{7}$ km. The 
action of the Fresnel filter for scale sizes $\geq$ $10^{3}$ km is such that it will 
give rise to scintillation at distances $>$ 1 AU or in other words the Earth would be well 
within the Fresnel or near zone for these scale sizes. The IPS phenomenon therefore 
has an in-built filter that makes it insensitive to contributions from large-scale 
size density irregularities. In fact this property of IPS has even been exploited to 
study the fine scale structure in cometary ion tails during radio source occultations by 
cometary tail plasma \citep{ARB75,JaA91,JaA92}.  

The degree to which compact, point-like, extragalactic radio sources exhibit scintillation,
as observed by ground-based radio telescopes, is quantified by the scintillation index (m) 
given by $\mathsf{m = \frac{{\Delta}S}{<S>}}$, where $\Delta$S is the scintillating 
flux and $<$S$>$ is the mean flux of the radio source being observed.  For a given 
IPS observation, m is simply the root mean-square deviation of the signal intensity 
to the mean signal intensity and can be easily determined from the observed intensity 
fluctuations of compact extragalactic radio sources.  

Though IPS measures only small scale fluctuations in density and not the bulk 
density itself, it has been shown \citep{HTG85} that there were no variations in 
IPS measurements of $\Delta{N}$ that were not associated with corresponding variations 
in density $N$. These authors used a normalized scintillation index `g' (a good proxy 
for the density) to derive a relation between `g', and the density given by 
g = ($N$ cm${^{-3}}$/9)${^{0.52 \pm 0.05}}$.

For an ideal point-like radio source and at an observing wavelength $\lambda$, 
m will steadily increase with decreasing distance `r' from the Sun until it reaches a 
value of unity at some distance from the Sun.  As r continues to decrease beyond this point, 
m will again drop off to values below unity.  This turnover distance is a function of 
observing frequency and at 327 MHz ($\lambda$ = 92 cm) occurs at $\approx$0.2 AU or 
$\approx$40R${_{\odot}}$. The region beyond the turn-over distance is known as the weak 
scattering regime.  In addition to the dependence on heliocentric distance, 
m will also reduce with an increase in the angular diameter of the radio source being observed.  
\subsection{Phase modulations of waves and Scintillation index}
 \label{S-Phase}

The assumption that the solar wind is considered to be a confined to a thin 
slab as depicted in Figure \ref{Fig1} is due to the fact that the 
solar wind scattering function $\beta$(r)~$\propto ~$r${^{-4}}$. Hence, most 
of the contribution to the scintillation will come from the point `P' on the LOS 
that is closest to the sun.  Plane waves from distant, compact extragalactic 
radio sources on passing through the thin slab of density irregularities will 
have an rms phase deviation ($\phi{_{rms}}$) imposed across their wave fronts.  The 
expression for $\phi{_{rms}}$ is
\begin{equation}
   \mathsf{{\phi}_{rms} = (2\pi){^{\frac{1}{4}}}~{\lambda}~{r_e}~(aL)^{\frac{1}{2}}~
   [<{{\Delta}{\it{N}}}^2>]^{\frac{1}{2}}}
   \label{eq2}
\end{equation}
where, r${_{e}}$ is the classical electron radius, $\lambda$ is the observing wavelength, 
and a is the typical scale size in the thin screen of thickness L (see Fig.\ref{Fig1}).  In 
the weak scattering regime, m is given by
\begin{equation}
     \mathsf{m \approx \sqrt{2}~\phi{_{rms}}}
     \label{eq3}
\end{equation}

Equations \ref{eq2} and \ref{eq3} can be rewritten as

\begin{equation}
   \mathsf{{\Delta}{\it{N}} = \frac{m}{(2){^{\frac{1}{2}}} (2\pi){^{\frac{1}{4}}} 
   {\lambda}{r_e}(aL)^{\frac{1}{2}} }}
   \label{eq4}
\end{equation}

Equation~(\ref{eq4}) gives us a prescription for determining the quantity 
$\Delta{N}$ from observations of $m$.

\section{Data analysis}
 \label{S-Data}
Regular IPS observations on a set of about 200 chosen extragalactic radio 
sources have been carried out to determine solar wind velocities and scintillation 
indices at 327 MHz \citep{KKa90,AsK98} since 1983 at the multi-station IPS 
observatory of STEL, Japan.  Prior to 1994, these observations were 
carried out by the three-station IPS facility at Toyokawa, Fuji, and Sugadaira.  
In 1994, one more antenna was commissioned at Kiso forming a four\,--\,station 
dedicated IPS network that has been making systematic and reliable estimates 
of solar wind velocities and scintillation indices \citep{TKF12} except for a data 
gap of one year in 1994.  Systematic observations have been carried out on about 
a dozen selected radio sources each day such that each source would have been observed 
over the whole range of heliocentric distances between 0.2 and 0.8 AU in a period of 
about 1 year.  We have employed the daily measurements of m, spanning the period 
from 1998 to 2008, covering solar cycle 23.

%
\protect\begin{figure}[htb]
\vspace{5.4cm}
\includegraphics{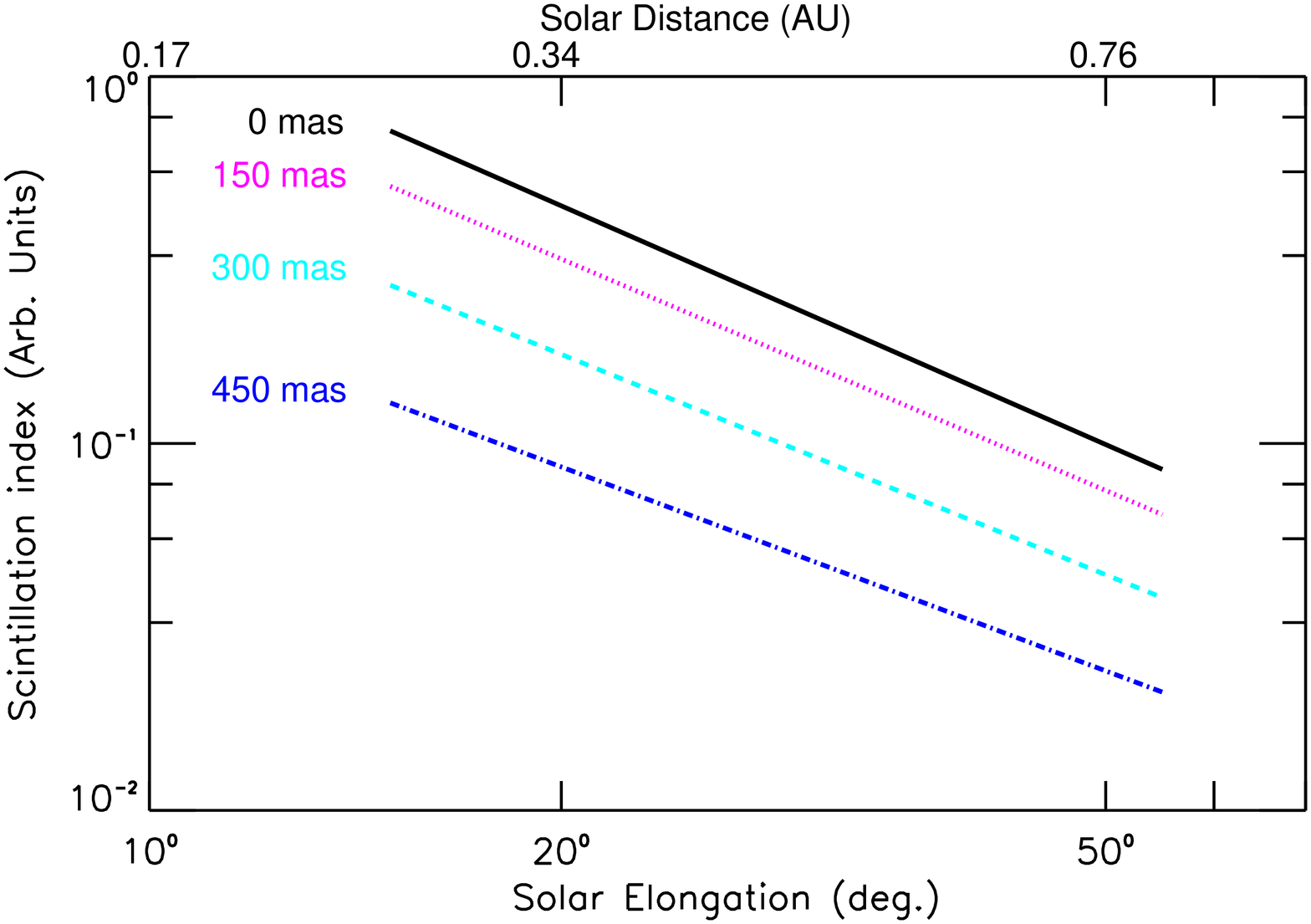}
\caption{shows curves of theoretically values of m as a function of solar elongation 
for various source sizes corresponding to sizes of 0 mas, 150 mas, 300 mas, and 450 
mas. These theoretical values of m are computed using \cite{Mar75} model.}
\label{Fig2} 
\end{figure}
%
Very compact radio sources are extremely rare and it has been established at a 
number of frequencies, using both IPS \citep{Bou69, BCr72, Mil76} and long baseline 
interferometry \citep{ClK68, ClB69} that the radio source 1148-001 has an angular 
diameter of $\approx$10 milli arcsecond (mas) at meter wavelengths.  Thus, the source 
1148-001 can be treated as a nearly ideal point source at 327 MHz, with almost 
all of its flux contained in a compact scintillating component with very little 
flux outside this compact component \citep{Swa77, VeA85}.  As stated earlier, for 
such ideal point sources m will be unity at the turn-over distance, and will drop as 
the distance of the LOS to the source moves further away from the sun. For sources 
with larger angular diameters, m will be less than unity at the turn-over distance. 

\cite{Mar75} computed values of m for radio sources of a given source size as a 
function of r by obtaining theoretical temporal power spectra using a standard solar 
wind model assuming weak scattering and a power law distribution of density 
irregularities in the IP medium.  Figure \ref{Fig2} shows curves of theoretical m, 
computed using the Marians model \citep{Mar75}, as a function of $\epsilon$ (in degrees) 
for source sizes of 0 mas, 150 mas, 300 mas, and 450 mas, respectively. All the 
curves are plotted for $\epsilon$ ranging from 15${^{\circ}}$ to 55${^{\circ}}$ 
corresponding to the weak scattering regime at 327 MHz which covers heliocentric 
distances between 0.26 and 0.82 AU.
%
\protect\begin{figure}[ht]
\vspace{7.6cm}
\includegraphics{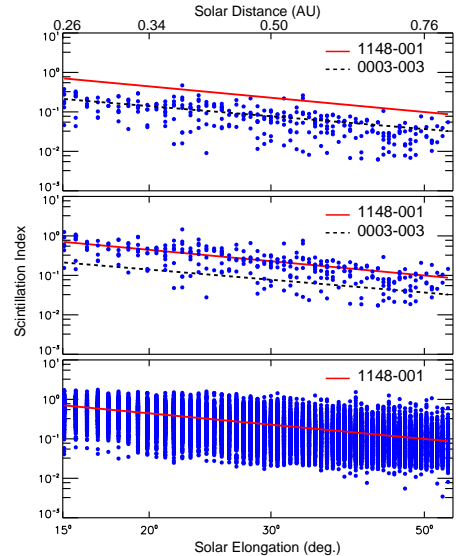}
\caption{The upper panel shows by filled blue dots, the actual measurements of normalized 
scintillation indices for the source 0003-003.  The theoretically computed curve for 
m using Marian's model \cite{Mar75} for both 0003-003 (dotted black) and 1148-001 (red line) 
are overplotted.  The middle panel shows the same two theoretical curves for sources, 
1148-001 and 0003-003 after the data of 0003-003 has been multiplied by a factor, 
determined from ratio of theoretical curves of 1148-001 and 0003-003 at each $\epsilon$, to 
remove the effects of source size. The lower panel shows the data for all 27 sources 
after being normalized to remove the source size effect. It can be seen that the data is 
well fitted to the theoretical curves of the source 1148-001.} 
\label{Fig3} 
\end{figure}
%
\protect\begin{figure*}[ht]
\vspace{9.0cm}
\includegraphics{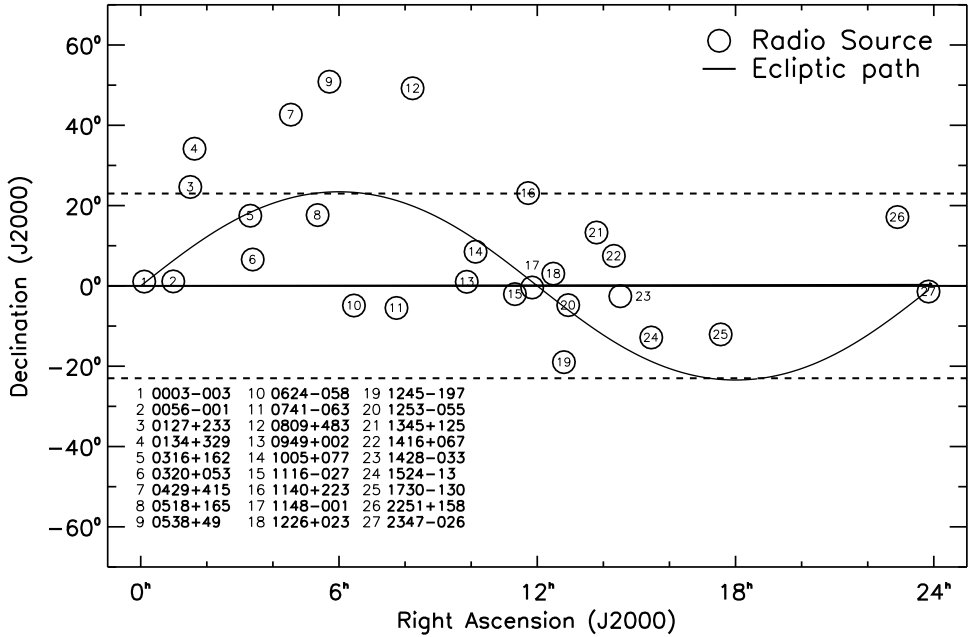}
\caption{Shows the coordinates (RA and Dec.) of the 27 selected radio sources by 
numbered open circles. The solid curve represents the path (RA and Dec.) of the Sun. 
Each numbered source name is indicated at the bottom left of the figure. }
\label{Fig4} 
\end{figure*}

For the present analysis and in order to obtain a uniform data set, it would be necessary to 
either choose sources of the same angular size or remove the effect of the finite source 
size by appropriately normalizing the data.  The normalization was carried out using a 
least squares minimization to determine which of the Marians curves best fits the data 
for a given source.  Since it is known that 1148-001 is a good approximation to a point 
source, the observed values of m of all other sources were multiplied by 
a factor equal to the difference between the best fit Marians curve for the given source 
and the best fit Marians curve for 1148-001, at the corresponding $\epsilon$. The best 
fit Marians curve for 1148-001 corresponds to that obtained for a source size of 10 mas.  

The upper panel of Figure \ref{Fig3} shows, by filled blue dots, one example of the 
actual observations of m as a function of heliocentric distance for the source 0003-003.  
The dashed red line is the Marians curve corresponding to a source size of 10 mas, while 
the dashed black line is the Marians curve which best fits the the data for the source 0003-003.  
The middle panel of Figure \ref{Fig3} shows the same data after it has been normalized, as 
described above to remove the effect of the finite source size.  After normalizing all 
the observations in the above manner, we shortlist only those sources which had at least 
400 observations distributed uniformly over the entire range of heliocentric distances 
without any significant data gaps.  Using this criteria we finally shortlisted 27 
sources for further analysis.  The normalized points for all 27 sources are shown 
in the lowermost panel of Figure \ref{Fig3} and they fit the theoretical curve of the source 
1148-001 very well. The Right Ascension and Declination (J2000 epoch) of 
the 27 shortlisted radio sources are shown in Figure \ref{Fig4} by numbered open circles with the 
corresponding names of the sources (B1950 epoch) listed at the bottom of Figure \ref{Fig4}.  The
ecliptic radio sources in Figure \ref{Fig4} are those in the declination range $\pm$23${^{\circ}}$, 
while the non-ecliptic or high latitude sources lie above this range of declinations. 

Using equation \ref{eq4}, ${\Delta}N$ has been obtained at heliocentric distances  
in the range 0.26\,--\,0.82 AU (55\,--\,175 $R_{\odot}$) from 1998 to 2008, using 
daily IPS measurements of m.  In order to estimate the background solar wind density, 
we use values of the daily average solar wind density ($N$) obtained from the Solar Wind 
Electron, Proton, and Alpha Monitor (SWEPAM) onboard the {\it{ACE}} spacecraft, covering 
the period from 1998 to 2008.  However, {\it{ACE}} density measurements are effectively 
at a distance of 1 AU. Thus, for estimation of density at the locations, spread over 
distances of 0.26\,--\,0.82 AU, the measured {\it{ACE}} densities at 1 AU were extrapolated 
in the sunward direction using a background density model by \cite{LDB98}.  According to 
this model, the background density, $N$ at r (in units of AU) is given by 
\begin{equation}
   \mathsf{ N = 7.2 r^{-2}+1.95 {\times}{10^{-3}}r^{-4}+8.1 {\times}{10^{-7}}r^{-6}~cm^{-3}}
   \label{eq5}
\end{equation}

This equation assumes a density of 7.2 $cm^{-3}$ at 1 AU. In order to derive the 
background density at a given $r$, we use equation~\ref{eq5} multiplied by $N$(1 AU)/7.2, 
where $N$(1 AU) denotes the value of the density from the {\it{ACE}} data.  As discussed 
earlier, the $\Delta N$ is deduced from IPS measurements of m using Eq.~\ref{eq4}.  We 
compute $N$ by using near-Earth {\it{ACE}} measurements that are contemporaneous with 
the measurement of $m$ and extrapolate it sunwards to the heliocentric distance where $m$ 
is measured.  For instance, let us consider the observation of the source 1148-001 in 1999 
at an $\epsilon$ (heliocentric distance) of 15$^{\circ}$(0.26 AU).  We use {\it{ACE}} data
at 1 AU from year 1999 and extrapolate it sunwards to a heliocentric distance of 0.26 AU to 
determine the appropriate N to be used in Eq~\ref{eqmodindex}.  The ratio of $\Delta N$ 
to $N$ gives the $\epsilon{_N}$ (Eq~\ref{eqmodindex}). As stated earlier, the m of a 
given source is a function of the both the distance of the LOS from the Sun and the source 
size, with ideal point-like radio sources giving an m of $\approx$1 at start of the weak 
scattering regime which, at 327 MHz, is at a distance of approximately 0.2 AU.  This is 
the reason that we can probe the solar wind at 327 MHz over a distance range of 0.26\,--\,
0.82 AU (55\,--\,175 $R_{\odot}$). 

\protect\begin{figure}[htb]
\vspace{9.0cm}
\includegraphics{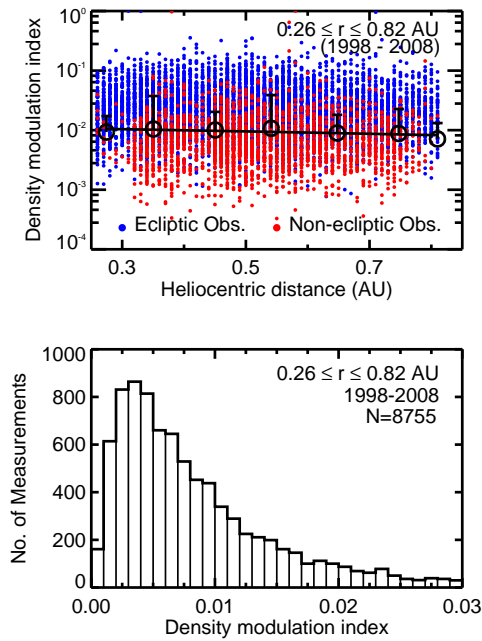}
\caption{shows, in the upper panel, spatial variation of the density modulation index, 
$\epsilon_{N}$, of all the 27 selected sources, in the period from 1998 to 2008. While 
the blue and red solid dots are the actual measurements of normalized modulation indices 
for ecliptic sources and non-ecliptic sources respectively, the large open circles in 
black represent averages of all observation at intervals of 0.1 AU.  The solid line is 
a fit to these average values.  The lower panel shows a histogram of the $\epsilon_{N}$, 
with a median and mean of 0.006 and 0.01 respectively.}
\label{Fig5} 
\end{figure}
%
\section{Temporal and Spatial Behavior of $\epsilon_{N}$}
 \label{S-mod-index}
The upper panel of Figure \ref{Fig5} shows the $\epsilon_{N}$ as function of r in the 
range 0.26 to 0.82 AU and spanning the period 1998\,--\,2008.  The solid blue and red 
dots represent the $\epsilon_{N}$ derived for ecliptic and non-ecliptic source 
observations respectively, while their running averages at heliocentric distance intervals 
of 0.1 AU are shown by large open circles with 1 $\sigma$ error bars.  The decline in 
the $\epsilon_{N}$ is only 0.22\%.  So it is quite apparent that $\epsilon_{N}$ is almost 
independent of heliocentric distance.  The solid black line is a fit to the running 
averages of $\epsilon_{N}$, which emphasizes this trend. The Marians model, by assuming 
a spherically symmetric distribution of density fluctuations ignores any  latitudinal 
structure in the density fluctuations.  IPS data of non-ecliptic sources are 
therefore likely to be affected by the latitudinal structure caused for example by polar 
coronal holes.  So, the difference between ecliptic and non-ecliptic sources may be 
attributed to a bias caused by the effect of the solar wind latitudinal structure. 

Histograms of $\epsilon_{N}$ for the 27 selected sources used in the present analysis 
are shown in the lower panel of Figure \ref{Fig5}.  The total number of measurements 
are mentioned on the top right corner of Figure \ref{Fig5}.  An inspection of the 
histogram of $\epsilon_{N}$ shows that $0.001 \lesssim \epsilon_{N} \lesssim 0.02$ with 
a most probable value of 0.006 and a mean of 0.01.  These values are somewhat lesser 
than the values of $0.03 \lesssim \epsilon_{N} \lesssim 0.08$ reported using {\it{Wind}} 
spacecraft measurements of density fluctuations at 1~AU \citep{SSp04}.  A modulation index 
$\epsilon_{N} \lesssim 0.1$ has been reported by \cite{BBr95} using measurements from 
the {\it{Helios 2}} spacecraft between 0.03\,--\,1~AU.  However, in both these papers, 
the data used covered only a limited time interval (albeit with a high sampling frequency 
of 45 min), whereas this study uses data for eleven years, covering almost the entire solar 
cycle 23 (with a sampling frequency of one day).  
\protect\begin{figure}[htb]
\vspace{8.75cm}
\includegraphics{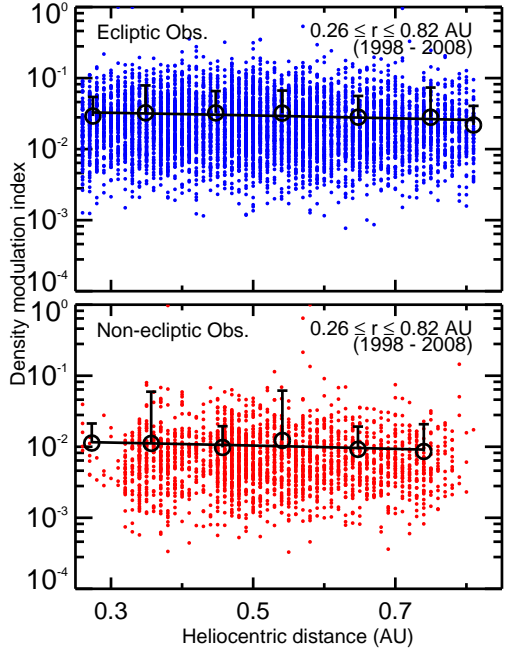}
\caption{shows, in the upper panel, spatial variation of the $\epsilon_{N}$ for  
ecliptic sources, in the period from 1998 to 2008.  While the lower panel shows the 
spatial variation of the $\epsilon_{N}$ of non-ecliptic sources.}
\label{Fig6} 
\end{figure}
%

Figure \ref{Fig6} shows the spatial variation of $\epsilon_{N}$ for IPS measurements of ecliptic 
(upper panel) and non-ecliptic sources (lower panel).   The mean values of $\epsilon_{N}$ for 
ecliptic and non-ecliptic sources are 0.03$\pm{0.03}$ and 0.01$\pm{0.02}$ respectively, 
showing a slightly higher $\epsilon_{N}$ for the ecliptic sources.  The decline in 
$\epsilon_{N}$ with heliocentric distance for the ecliptic and non-ecliptic sources are 0.7\% 
and 0.25\% respectively.  So it is again clearly evident that $\epsilon_{N}$ is independent 
of heliocentric distance for both ecliptic and non-ecliptic sources.

\subsection{Long-term Temporal Changes of $\epsilon_{N}$}
 \label{S-tempo-mod-index}
%
%
\protect\begin{figure}[ht]
\vspace{5.00cm}
\center
\includegraphics{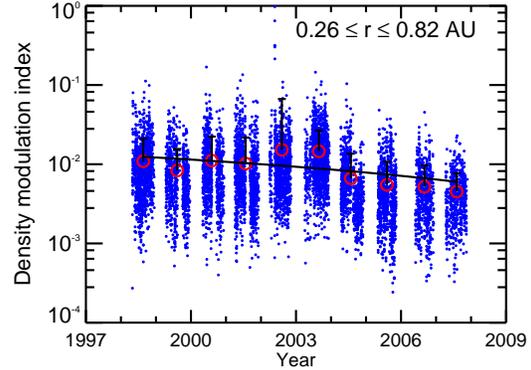}
\caption{shows the $\epsilon_{N}$ as function of time for the selected 27 sources, 
at heliocentric distances of 0.26$-$0.82 AU.  While the blue solid dots are the actual 
measurements of $\epsilon_{N}$, the large open circles in red represent annual means.  
The solid curve is a linear fit to annual means of $\epsilon_{N}$.}
\label{Fig7} 
\end{figure}
%
A study of the long-term changes in IPS measurements of m, a good proxy for solar 
wind microturbulence levels, has shown a systematic and steady decline in m since 
$\approx$1995 \citep{JaB11}.  One would therefore expect that electron density 
fluctuations, ${\Delta}N$ would also exhibit a similar decrease. In fact, a 
consistent decrease in electron density turbulence, in regions of the inner-heliosphere 
has been reported \citep{TKF12} using IPS measurements from STEL.  Using IPS measurements 
from the Ooty Radio Telescope \cite{Man12} also reported a declining trend of the density 
turbulence from the year 2004 to 2009 (see Figure 3 in \cite{Man12}).  It is therefore 
of interest to see how the $\epsilon_{N}$ during the period 1998\,--\,2008 vary in time.  

Figure \ref{Fig7} shows the temporal variation of $\epsilon_{N}$, covering the 
period 1998\,--\,2008, at heliocentric distances ranging from 0.26 to 0.82 AU.  
The blue solid dots are the derived density modulation indices while annual 
means of the modulation indices are shown by large red open circles with 1 sigma 
error bars. The annual means of $\epsilon_{N}$ show a decline of 8\% in 
$\epsilon_{N}$.  This finding impacts our understanding of the steady temporal 
decline in solar wind dynamic pressure; we discuss this further in the next 
section.
\section{Summary}
 \label{S-dis}
\subsection{Conclusions}
We have carried out an extensive survey of the density modulation index ($\epsilon_{N}$) 
in the inner-heliosphere using IPS observations at 327 MHz.  We have used observations of 
27 sources spanning the heliocentric distance range 0.26\,--\,0.82 AU for the period 
1998\,--\,2008. One of the broad conclusions of our study is that $\epsilon_{N} \approx 
0.01$, and is roughly constant with heliocentric distance.  Our result shows the typical 
amplitudes of density modulation index are low, of the order of 0.1\%\,--\,2\% and these 
values are somewhat lower than the values of 3\%\,--\,8\% reported by \cite{SSp04}.  It may 
be noted, however, that \cite{SSp04} have used only near-Earth observations, whereas our 
observations span a heliocentric distance range of 0.26\,--\,0.82 AU.  Earlier measurements 
(\cite{TMa94} and \cite{ BBr95}) of $\epsilon_{N}$ from Helios data at heliocentric distances 
between 0.3 and 0.5~AU have found 5\% $\lesssim \epsilon_{N} \lesssim$ 20\%.  

Our result of $\epsilon_{N}$ being independent of heliocentric distance agrees with those 
proposed by \cite{WoA95} for the slow solar wind.  Using {\it{Ulysses}} time delay measurements, 
\cite{WoA95} have shown that the relative density fluctuations obtained over a period of 
5 hours for the slow solar wind ($\leq$250 km s${^{-1}}$) in the distance range from 0.03 to 
1 AU is independent of heliocentric distance.

The long-term temporal variation of the relative density fluctuations over heliocentric 
distances of 0.26\,--\,0.82~AU, have shown a decline of 8\% during the period 1998\,--\,2008.  

\subsection{Discussion}

We now comment on the implications of our results on some of the problems we have outlined in 
the introduction:

\protect\begin{figure*}[ht]
\vspace{8.00cm}
\includegraphics{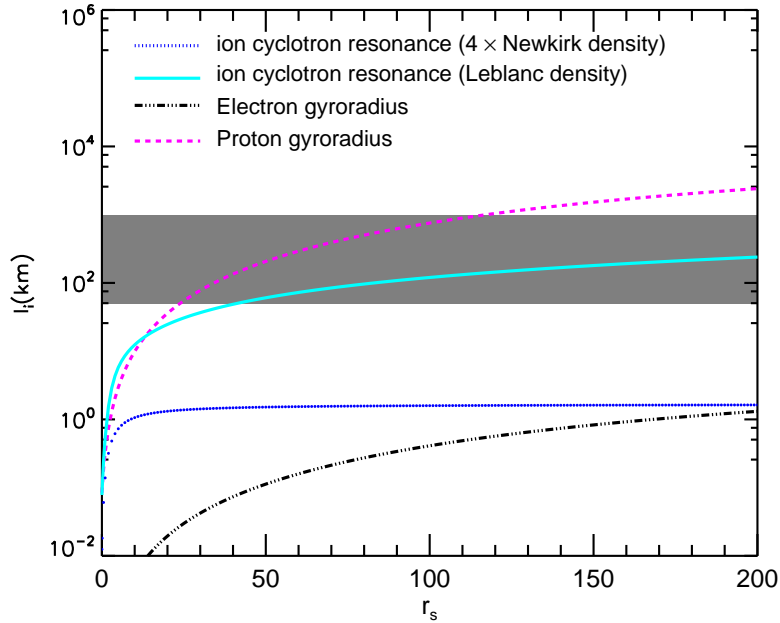}
\caption{The inner scale $l_i$ in km as a function of heliocentric distance in units of solar 
radii ($r_s$). The dashed lines show the proton gyroradius using a proton temperature of $10^5$~K. 
The solid and dotted lines shows the inner scale governed by ion cyclotron resonance using 
the Leblanc et al density model and the fourfold Newkirk density model respectively. The 
dot-dashed line shows the electron gyroradius using an electron temperature of $10^5$~K. The 
light gray region denotes the range of spatial scales for which IPS observations are sensitive.}
\label{Fig8} 
\end{figure*}
\begin{itemize}
\item
The scintillation levels in the inner-heliosphere (which are $\propto \Delta$N) have 
been shown to be declining monotonically since $\approx$ 1995 \citep{JaB11,TKF12}.  
Assuming that $\Delta N \propto$ the background density N, this has prompted speculations 
about a steady temporal decline in the pressure exerted by the bulk solar wind on the 
Earth's magnetosphere.  \cite{McA13} have calculated the canonical standoff 
distance of bow shock nose of the Earth's magnetosphere which is about 11 Earth radii 
($R_{E}$) for the period 2009\,--\,2013 compared to about 10 $R_{E}$ for the period 
1974\,--\,1994.  According to these authors, this change is in view of the observed 
decline in solar wind dynamic pressure from $\sim$2.4 nPa (1974\,--\,1994) to $\sim$1.4 
nPa (2009\,--\,2013).  However, these need to be revisited in light of our findings of a 
small, but discernible, steady decrease in $\epsilon_{N} \equiv \Delta N/N$ with time.

Furthermore, if there is a linear relationship between the relative density fluctuations 
and the magnetic field fluctuations \citep{SSp04}, it would imply that the magnetic field 
fluctuations also decline steadily over period 1998\,--\,2008.  So it appears reasonable 
to conclude that the decrease in density fluctuations is connected to the unusual 
solar magnetic activity during the long deep solar minimum at the end of the solar cycle 
23.  It has been shown that both solar polar fields and the level of turbulent density 
fluctuations ($\Delta N$) have decreased monotonically since around 1995 \citep{JBG10,
JaB11,BiJ14}. 

\item
We note that the IPS technique used in this work to infer density fluctuations is 
sensitive to spatial scales of 50 to 1000 km \citep{RBA74,CFi85,FBD08}.  It is worth 
examining how these scales relate to the dissipation scale of the turbulent cascade 
(often referred to as the inner scale).  If the length scales probed by the IPS technique 
are in the inertial range, it is reasonable to presume that the magnetic field is frozen-in, 
and the density fluctuations can then be taken as a proxy for magnetic field fluctuations 
(e.g., \cite{Spa02}).  We note, however, that the flux-freezing concept might not hold 
for turbulent fluids (e.g., \cite{LVi99}).  In order to investigate this issue, 
we consider three popular inner scale prescriptions.  One prescription for the inner scale 
assumes that the turbulent wave spectrum is dissipated due to ion cyclotron resonance, and 
the inner scale is the ion inertial scale \citep{CHa89}.  In this case, the inner scale 
($l_{i}$) is given as a function of heliocentric distance $r$ by
\begin{equation}
l_i® = 684 \, n_e(r)^{-1/2} \,\,\,{\rm km}
\label{inertial}
\end{equation}
where $n_{e}$ is the number density in ${\rm cm}^{-3}$.  A second prescription identifies the 
inner scale with the proton gyroradius \citep{BaK05,AlL12}.  In this case the inner scale is 
given by
\begin{equation}
l_{i}(r) =  1.02\times 10^2 \mu^{1/2} T_i^{1/2} B(r)^{-1}  \,\,{\rm cm}
\label{protonscale}
\end{equation}
where $\mu  (\equiv m_p/m_e)$ is the proton to electron mass ratio, $T_i$ is the proton 
temperature in eV and B is the Parker spiral magnetic field in the ecliptic plane \citep{Wil95}. 
However, recent work seems to suggest that the dissipation could occur at scales as small as 
the electron gyroradius \citep{AlL12,SaH13}. The third prescription we therefore consider is 
one where the inner scale is taken to be equal to the electron gyroradius and is given by
\begin{equation}
l_{i}(r) = 2.38\times T_e^{1/2} B(r)^{-1}  \,\,{\rm cm}
\label{electronscale}
\end{equation}
where $T_e$ is the electron temperature in eV. The inner scales using these three prescriptions 
(Eqs~\ref{inertial}, \ref{protonscale}, \ref{electronscale}) are shown in Figure~\ref{Fig8} as 
a function of heliocentric distance. The grey band denotes the range of length scales ($\approx$ 
50 - 1000 km) to which the IPS technique is sensitive. As explained in the caption of Figure 
\ref{Fig8}, we use electron and proton temperatures of $10^5$K in order to compute the proton 
and electron gyro radii respectively.  The magnetic field is taken to be a standard Parker spiral 
\citep{Wil95}.  In order to compute the inner scale using Eq~(\ref{inertial}), we need a density model.  
We have used two representative density models -- the Leblanc density models \citep{LDB98} and the 
fourfold Newkirk density model \citep{New61}.  If the length scales probed by the IPS technique 
(denoted by the grey band in Figure~\ref{Fig8}) are larger than the inner scale, we can conclude 
that the density fluctuations discussed in this paper lie in the inertial range of the turbulent 
spectrum.  From Figure~\ref{Fig8}, it is evident that this is the case all the way from the Sun 
to the Earth only if the inner scale is the electron gyroradius, or if it is due to proton cyclotron 
resonance, and the density is given by the fourfold Newkirk model. On the other hand, if the inner 
scale is given by the proton gyroradius, or if the inner scale is due to proton cyclotron resonance 
and the density model is given by the \cite{LDB98} prescription, the density fluctuations probed by 
the IPS technique are probably smaller than the dissipation scale for heliocentric distances 
beyond 30--40 $R_{\odot}$.

\item
In order to account for the magnitude of cosmic ray Forbush decreases observed at the 
Earth, \cite{SuA09} and \cite{ArA13} deduce that the level of magnetic field turbulence 
in the sheath region ahead of Earth-directed CMEs ranges from a few to a few 10's of 
percent.  The magnetic field turbulence level is often taken to be a proxy for $\epsilon_{N}$ 
\citep{Spa02}.  Generally, the turbulence level in the sheath region would be expected 
to be somewhat higher than (but not very different from) its value in the quiescent solar 
wind.  The results of this paper regarding the magnitude of $\epsilon_{N}$ in the quiescent 
solar wind are thus broadly consistent with the deductions of \cite{SuA09} and \cite{ArA13} 
regarding  the magnetic field turbulence level.

\item
Reid \& Kontar (2010) have argued that the modulation index $\epsilon_{N}$ 
needs to be around 10\% near the Earth and be proportional to $R{^{0.25}}$ 
(where $R$ is the heliocentric distance) in order to account for the 
Earthward transport of electron beams produced in solar flares.  However, 
we find that the modulation index shows no change with increasing heliocentric 
distance, and that its value near the Earth is considerably smaller than 10\%.

\end{itemize}

\section{Acknowledgments}
IPS observations were carried out under the solar wind program of STEL, Japan. 
We thank the ACE SWEPAM instrument team and the ACE Science Center for 
providing the ACE data available in the public domain via World Wide Web.

\bibliographystyle{apj}

\clearpage

\end{document}